\documentclass{ws-procs9x6}
\usepackage{graphicx}
\usepackage{epsfig}
\newcommand{\GeVcc}    {\mbox{$ {\mathrm{GeV}}/c^2                           $}}

\newcommand{\hetrois}    {\mbox{$ ^{3}{\mathrm{He}}                            $}~}
\newcommand{\hetro}    {\mbox{$ ^{3}{\mathrm{He}}                            $}}

\newcommand{\neut}{$\tilde{\chi}$~}
\newcommand{\neutt}{$\tilde{\chi}$}

\newcommand{\gam} {{$\gamma$-ray}~} 
\newcommand{\gams} {{$\gamma$-rays}~}

\def\NIMA#1#2#3{{\rm Nucl.~Instr.~and~Meth.} {\bf{A#1}} (#2) #3}

\def\PLB{{\em Phys. Lett.}  B}

\def\PRB#1#2#3{{\rm Phys. Rev.} {\bf{B#1}} (#2) #3}

\def\PRL#1#2#3{{\rm Phys.~Rev.~Lett.} {\bf{#1}} (#2) #3}
\def\PLB#1#2#3{{\rm Phys.~Lett.} {\bf{B#1}} (#2) #3}

\begin{document}
%\runauthor{}
%{\flushleft {\bf Note interne ISN :} ISN-00116}\\
%\begin{frontmatter}
\title{MACHe3~: A new generation detector for non-baryonic dark matter direct detection}
\author{D.~SANTOS~$^{1}$, F.~MAYET~$^{1}$, E.~MOULIN~$^{1}$, G.~PERRIN~$^{1}$, Yu.~M.~BUNKOV~$^{2}$, 
H.~GODFRIN~$^{2}$, M.~KRUSIUS~$^{3}$}
\address{$^{1}$ Institut des Sciences Nucl\'eaires, 
 CNRS/IN2P3 and Universit\'e Joseph Fourier, 
 53, avenue des Martyrs, 38026 Grenoble cedex, France}
\address{$^{2}$ Centre de Recherches sur les Tr\`es Basses Temp\'eratures,
 CNRS and Universit\'e Joseph Fourier, BP166, 38042 Grenoble cedex 9, France} 
\address{$^{3}$ Low Temperature Laboratory, Helsinski University of Technology, Finland}
\maketitle
\abstracts{
MACHe3 (MAtrix of Cells of superfluid \hetro) is a project of a new detector for direct Dark Matter (DM) search, 
using superfluid \hetrois as a sensitive medium. 
An experiment on a prototype cell has been performed and the first results reported here 
are encouraging to develop of a multicell prototype.
In order to investigate the discovery potential of MACHe3, and its
complementarity with other DM detectors, a phenomenological study done 
with the DarkSUSY code is shown.}
%
%
%
%--------------------------------------------------------------------
%
\section{Introduction}
In the last decades many promising detectors have been developed to search for non-baryonic dark matter.
Recent results from these new detectors have improved the upper limits on scalar and 
axial interaction exclusion plots\cite{excledel}. 
The main difficulties for direct detection techniques concern neutron
interactions and intrinsic impurities of the sensitive medium and surrounding materials.
Following early experimental works\cite{lanc}, a superfluid \hetrois detector has 
 been proposed\cite{firstmac3} for direct Dark Matter search. Monte Carlo simulations have 
shown that a high granularity detector, a matrix of superfluid \hetrois cells, would allow to reach a high rejection factor
against background events, leading to a low false event rate. 
\subsection{Superfluid \hetro-B as a privileged sensitive medium }

The use of the superfluid \hetrois in the B phase as a sensitive medium for direct detection has the following advantages 
with respect to other materials~:
\begin{enumerate}
\item \hetrois being a 1/2 spin nucleus, an \hetrois detector will be mainly sensitive  
to axial interaction, making this device complementary to existing ones. The axial
interaction is largely dominant in all the SUSY region associated 
with a substantial elastic cross-section\cite{thesefmayet}.

\item  Its low mass allows a wider range of sensitivity to the WIMP mass.

\item  Low Compton and photoelectric cross-sections. No intrinsic X-rays.

\item  A high neutron capture cross-section, producing a clear signature well discriminated from a WIMP signal.

\item  A low threshold, less than 1 keV.

\item  An ultra-high purity due to its superfluid phase.

\item  A high signal to noise ratio, due to the small energy range expected for a WIMP signal.
The maximum recoil energy does only slightly depend on the
WIMP mass, due to the fact that the target nucleus (${\rm m=2.81 \, GeV}\!/c^2$) is much
lighter than the incoming  \neut (${\rm M_\chi \geq 32 \,GeV}\!/c^2$). As a matter of fact, the recoil
energy range needs to be studied only below ${\rm 6 \,keV}$, see\cite{firstmac3,thesefmayet}. 
\end{enumerate}
\subsection{Experimental device}

\begin{figure}[!ht]
\begin{center}
\epsfxsize=14pc   %width of figure - will enlarge/reduce the figures
\epsfbox{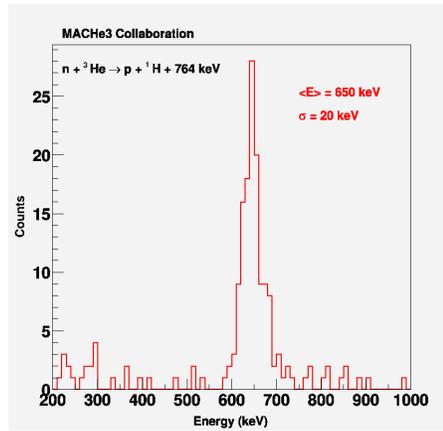}    
\caption{Neutron capture spectrum of the \hetrois-cell bolometer.}
\label{fig:neutroncapture}
\end{center}
\end{figure}

The elementary component of MACHe3 is the superfluid \hetrois cell\cite{prb98}. It is a small copper cubic box 
($\mathrm{V} \simeq 125 \,{\rm mm}^3$) filled with superfluid \hetrois in the B phase. This ultra low temperature device
 (T $\simeq \!100 \,\mu\mathrm{K}$) presents a low detection 
threshold ($\mathrm{E}_{th}\simeq 1 \,\mathrm{keV}$). An experimental test of such a prototype cell has been done at 
CRTBT in June 2001 with a lead shielding around the cryostat to estimate the contribution of the external
natural radioactivity. An experiment with an Am-Be neutron source has been performed to validate the response of the
cell. The n-capture peak and n-elastic collision events at low energies 
are shown on fig.~\ref{fig:neutroncapture}. 
The peak at 650 keV shows that a 15 \% of the total energy released by the capture (764 keV) is lost in scintillation and vortex production.
To disentangle these effects we are planning to develop a new prototype cell with an additional signal from the
temperature change of the cell walls produced by the absorption of scintillation UV photons. 
Preliminary results\cite{thesefmayet,santosgamma} also show that a threshold value down to $\sim 1 \,{\rm keV}$ 
has been achieved, and that a stability of the order of one week per cycle, at T $\simeq \!100 \,\mu\mathrm{K}$, has been obtained.

The final version of the detector will be a matrix of 1000 cells of $125 \,{\rm cm}^3$ each. The idea is to 
take advantage both on the 
energy loss measurement and the correlation among the cells to discriminate neutralino events from background ones 
(neutrons, \gams and muons). The design of the matrix has been optimized with a Monte Carlo 
simulation\cite{firstmac3}. 
In the preferred configuration, a $10 \,{\rm kg}$ detector, the false event rate has been shown to be as small as 
$\sim\! 10^{-1}\, {\rm day}^{-1}$ for neutron events and $\sim\! 10^{-2}\, {\rm day}^{-1}$ for muon events\cite{firstmac3}. In addition, background from \gam events needs to be 
taken into account. Energy loss measurement and correlation among the cells within a $10 \,{\rm kg}$ detector 
allows to obtain a rejection up to 99.8\% for $\sim 2\,{\rm MeV}$ $\gamma$-rays\cite{firstmac3}. Additional internal tag 
on $\gamma$-rays may be obtained using 
a new matrix configuration 
in which two neighbouring cells share a common copper wall, thus greatly improving correlation factor while reducing the amount
of copper used. Monte Carlo studies are under way\cite{santosgamma}. 

\section{Phenomenological study}
\label{sec:xsrate}
A phenomenological study has been done with the DarkSUSY code\footnote{The version used is 3.14.01, 
with correction of some minor bugs.}\cite{ds}, within the framework of the 
phenomenological Supersymmetric Model.
All the free parameters have been scanned on a large range\cite{plbf}.
This scan corresponds to a total number of supersymmetric (SUSY) models of the order of $2 \times 10^6$.  
In order to exclude SUSY models giving a \neut relic density too 
far away from  the estimated matter density in the Universe ($\Omega_{\mathrm{M}} \simeq 0.3$), only 
models with $\Omega_\chi$ in the following range are considered~:
$0.025 \leq \Omega_\chi \mathrm{h_0}^2 \leq 1$,
where $\mathrm{h_0} = (0.71 \pm 0.07)\times^{1.15}_{0.95}$ is the normalized Hubble expansion 
rate\cite{pdg2000}.
Standard galactic halo parameters have been used, in a spherical isothermal distribution, 
with a local density ($\rho_0 \!=\!   0.3\,{\rm GeV}\!/c^2\,{\rm cm}^{-3}$) 
and an average velocity~($v_0 \!=\! 220\,{\rm km}\,{\rm
s}^{-1}$).

In the general WIMP case, the allowed interactions are vectorial,
axial and scalar. The neutralino being a Majorana fermion, the vectorial interaction vanishes, leaving two
classes of interaction~: scalar (spin independent) and axial (spin dependent), the latter 
obviously requiring a non-zero spin nucleus. 
\hetrois being a light 1/2 spin nucleus, a medium made of such nuclei will be sensitive mainly to axial 
interaction\cite{thesefmayet}. 

Using the DarkSUSY code, the \neutt-\hetrois axial
cross-section has been evaluated. The calculation of the 
\neutt-quark elastic scattering amplitude is done at the tree-level, via an exchange of squark or $Z^0$. 
It has been shown\cite{plbf} that, for SUSY models not excluded by collider 
experiments and giving $\Omega_\chi$ within the range of interest, a cross-section as
high as $\sim \!10^{-2} \,{\rm pb}$ can be obtained for a $\sim \!60\,$\GeVcc~neutralino. 
The \neut event rate has then been evaluated for a $10\,{\rm kg}$ \hetrois matrix, 
to be compared with the
background rate\cite{firstmac3}.
A large number of models are giving a rate higher than the estimated false event rate induced by neutrons 
($\sim 10^{-1}\;\mathrm{day}^{-1}$), or above the estimated muon background ($\sim 10^{-2}\;\mathrm{day}^{-1}$).
It can be concluded that a high granularity \hetrois detector would present a sensitivity to a large part of
the SUSY region. The $\mu$ background level ($10^{-2}\;\mathrm{day}^{-1}$) is chosen as the lowest reachable limit
for MACHe3 and is taken hereafter as the reference value. 
\begin{figure}[ht]
\begin{center}
\epsfxsize=14pc 
\epsfbox{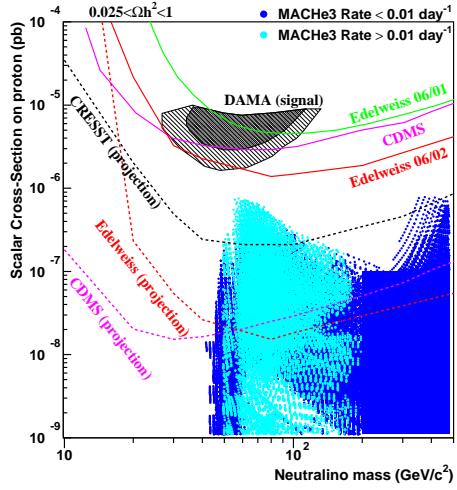}
\caption{Scalar cross-section (on proton) as a function of the \neut mass. 
Exclusion limits from the Edelweiss and CDMS experiments are 
shown, as well as the $3\,\sigma$ DAMA region.  
Dotted lines indicate projected limits from CRESST, Edelweiss and CDMS. 
Light (resp. dark) points indicate SUSY models giving a \neut rate in MACHe3 higher (resp. lower) 
than the estimated background level of $10^{-2}\,{\rm day}^{-1}$.}
\label{fig:damamac3}
\end{center}
\end{figure}
%
%-------------------------------------------------------------------------
%
\section{Complementarity with existing devices}
MACHe3 discovery potential may then be compared with direct detection experiments. 
It is indeed worth understanding whether planned projects would be sensitive to different SUSY regions.  
The correlation between axial and scalar detection have thus been investigated.
Many detectors for scalar interaction are already running or planned in the near future. 
At first sight, these two direct searches should be largely
independent as they involve different processes, e.g. Feynman diagrams 
at the tree level are different. 

Within the framework of the study described above, the proton scalar cross-section has been 
evaluated at the tree-level. Figure~\ref{fig:damamac3} presents the scalar cross-section on proton as a 
function of the \neut mass, for all models compatible with collider experiments and cosmology constraints. 
It can be seen that many SUSY models lie below the projected limits of future scalar detectors 
(Edelweiss\cite{excledel}, CDMS\cite{exclcdms}, CRESST\cite{projcresst}), while giving an event rate above the value taken as the lowest reachable limit for MACHe3.

\begin{figure}[t]
\begin{center}
\epsfxsize=19pc 
\epsfbox{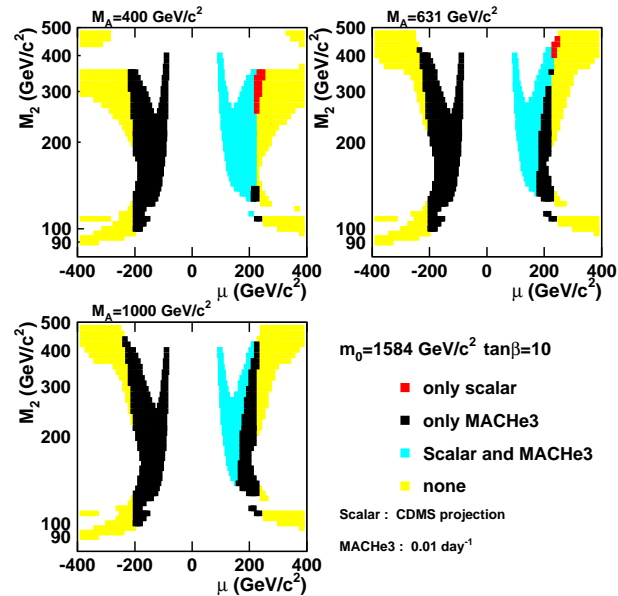}
\caption{Sensitivity regions for scalar search and MACHe3, in the Gaugino (${\rm M_2}$) Higgsino (${\mu}$) mass parameter 
plane, for $\tan \beta = 10$, ${\rm m_0 \simeq \;1.6 \,TeV\!/}c^2$ and 
three values of ${\rm M_A}$. The central white region is excluded by collider limits, while the two
other white regions are giving a relic density outside the range of interest. The color code
indicates the region potentially covered by MACHe3, 
by scalar detection, by both
strategies, or none of them.}
\label{fig:compdamamac3}
\end{center}
\end{figure}

As an illustration of this complementarity, we present on fig.~\ref{fig:compdamamac3} the regions potentially
covered by direct scalar search and by MACHe3, in a given part of the SUSY parameter space. The reference
limits have been chosen as~: the CDMS projected limit\cite{rick} and the $10^{-2}\,{\rm day}^{-1}$ limit for
MACHe3. 
It can be noticed that only axial direct search may be sensitive to the $\mu\!<\!0$ region, in  this given 
part of the SUSY parameter space, due to the fact that scalar interaction vanishes. 
\section{Conclusion}
It has been shown that a ${\rm 10\,kg}$ high granularity \hetrois detector (MACHe3) would allow to obtain, in many SUSY models, 
a \neut event rate higher than the estimated background. MACHe3 would thus 
potentially allow to reach a large part of the SUSY region, not excluded by current collider limits and for 
which the neutralino relic density lies within the range of interest. Furthermore, it has been shown that this project of new
detector would be sensitive to SUSY regions not covered by future or ongoing direct DM search detectors.

\end{document}